\documentclass{paper}
\usepackage{latexsym}
\usepackage{graphicx}
\usepackage{mathptmx}
\usepackage{algorithmic}
\usepackage{algorithm}
\usepackage{amsmath}
\usepackage{amsfonts}
\usepackage{amssymb}
\usepackage{amsbsy}
\usepackage{amsthm}

\usepackage[pdftex,colorlinks=true,urlcolor=blue,citecolor=black,anchorcolor=black,linkcolor=black]{hyperref}

\newtheoremstyle{wsc}
{3pt}
{3pt}
{}
{}
{\bf}
{}
{.5em}
{}

\theoremstyle{wsc}
\newtheorem{theorem}{Theorem}

\newcommand{\pr}{\mathbb{P}}

\begin{document}

%
%

\title{RANDOM GENERATION OF COMBINATORIAL STRUCTURES:\\ BOLTZMANN SAMPLERS AND BEYOND}

\author{Philippe Duchon\\ \\
        LaBRI, INRIA/Universit\'e de Bordeaux\\
        351, cours de la Lib\'eration\\
        33405 Talence cedex, FRANCE}

\maketitle

\begin{center}
This paper is dedicated to Philippe Flajolet (1948-2011) -- an
 outstanding mathematician\\ and computer scientist, and a great person
 to work with.  \end{center}

\section*{ABSTRACT}
The Boltzmann model for the random generation of ``decomposable''
combinatorial structures is a set of techniques that allows for
efficient random sampling algorithms for a large class of families of
discrete objects. The usual requirement of sampling uniformly from the
set of objects of a given size is somehow relaxed, though uniformity
among objects of each size is still ensured. Generating functions,
rather than the enumeration sequences they are based on, are the
crucial ingredient.

We give a brief description of the general theory, as well as a number
of newer developments.

\section{INTRODUCTION}
\label{sec:intro}

\subsection{Random Generation of Combinatorial Structures: a Quick Overview}

Random generation is often used as a tool for exploration (``what do
large objects of this type look like?''), or to provide large datasets
for software and algorithm testing. 

Usually, one defines a combinatorial class as a finite or countable
family $\mathcal{C}$ of discrete ``objects'', equipped with a ``size''
function from $\mathcal{C}$ to the natural numbers, such that for each
natural number $n$, the subset $\mathcal{C}_n$ of all objects with
size $n$ is finite. A \emph{uniform} random generator for the class
$\mathcal{C}$ is then a randomized algorithm that takes as input an
integer $n$, and outputs an element randomly selected uniformly from
$\mathcal{C}_n$.  The efficiency of the generator is typically
measured in terms of its expected time and space complexities,
expressed as a function of the size $n$.

Random generation methods tend to fall into one of a small number of
classes. \emph{Ad hoc} methods rely on precise combinatorial
properties of the considered class; a fine example is provided by
R\'emy's algorithm~\cite{Re85} for the random generation of plane
binary trees. \emph{Markov chain} methods rely on the simulation of a
Markov chain whose states are the objects one wants to sample from,
and that converges to the uniform (or another suitably chosen)
distribution; careful analysis of the convergence speed makes it
possible to run the chain for a fixed number of steps and obtain an
almost-uniform generator, or more sophisticated techniques such as
Coupling from the Past~\cite{ProppWilsonCFTP} can be used for exact
sampling from the stationary distribution.

\emph{Decomposition} methods apply for classes where objects are, 
informally, ``made up'' of smaller objects, be they from the same or
another class that is itself decomposable. A prime example is that
of \emph{plane binary trees}, that is, rooted trees where each
internal node has exactly two children, one of which is distinguished
as the \emph{left} child while the other is the \emph{right} child. In
this case, a plane binary tree is either made up of a single
root-leaf, or of a root and left and right subtrees, both of which can
be any plane binary trees.

The first systematic example of a decomposition method for random
generation is the so-called \emph{recursive method}~\cite{FlZiVC94},
where decompositions are used to obtain recurrences satisfied by the
counting sequences. These counting sequences are then used, together
with the decomposition rules themselves, to guide the random
generation algorithm. In the above example of plane binary trees, the
counting sequence (the well-known Catalan numbers) is used to
determine the probability $p_{n,k}$ that a uniform random tree with
$n$ internal nodes has a left subtree with $k$ internal nodes, and to
sample $K$ from this distribution; then the random generator is
recursively called with sizes $K$ and $n-1-K$ to obtain the left and
right subtrees (which are independent conditioned on their respective
sizes); the resulting tree is then uniform.

\subsection{The Boltzmann Method}

The \emph{Boltzmann method}, as introduced
in~\cite{DuFlLoSc02,DuFlLoSc04}, is another decomposition-based method
that can be applied to roughly the same combinatorial classes as the
recursive method. We will give a precise description of the method in
the next section. For this introduction, we will simply describe the
crucial ingredients.

Where the recursive method uses counting sequences for the random
generation algorithms, and generating functions are mostly a tool to
compute the counting sequences, the Boltzmann method uses the
generating functions, viewed as analytic functions of a real variable
(and, in practice, the values of the generating functions) in the
random generation algorithms -- thus reducing the need for
precomputation to a small number of real constants. The idea is to
``relax'' the requirement for a uniform sampler (which outputs a
uniform random structure among those of the target size $n$) into
allowing structures of all sizes, while keeping uniformity among all
objects of each individual size. By choosing the ``right''
distribution on sizes, independence among substructures is introduced,
which results in very simple and efficient algorithms.

\subsection{Outline of the Paper}

Sections \ref{sec:boltzmann}-\ref{sec:complexities} make up the bulk
of what can be termed the ``Boltzmann
method''. Section~\ref{sec:boltzmann} gives a general description of
the Boltzmann method. Section~\ref{sec:constructions} lists a number
of constructions which can be used to define specifications of
combinatorial classes for which Boltzmann samplers can be
automatically compiled from the
specification. Section~\ref{sec:complexities} sums up various results
on the complexities of the random sampling algorithms.
Section~\ref{sec:size} describes how Boltzmann samplers can be used to
get closer to the classical model of uniform, fixed-size random
generation.

Section~\ref{sec:gf_eval} deals with the question of how one can
effectively obtain the real constants used in Boltzmann samplers, and
Section~\ref{sec:not-quite} describes results that draw on the
principles of the method without exactly fitting in it.

The list of references does not attempt to give a complete list of
articles using the ideas exposed here. The interested reader will find
more examples in the bibliography of~\cite{Bo10}, even though it is
not limited to references about the Boltzmann method.

All proofs and most technical details have been purposefully omitted;
the algorithms in Section~\ref{sec:constructions} have been included
mostly to demonstrate their simplicity. We have made the choice of not
detailing any of the many examples that could be given; the interested
reader will find many such examples, including pictures of large
random structures, in the original papers.

\section{THE BOLTZMANN METHOD: GENERAL DESCRIPTION}
\label{sec:boltzmann}

The Boltzmann method can be used in two flavors, the \emph{ordinary}
(or unlabelled) and \emph{exponential} (labelled) variants.

Throughout the paper, we use the word \emph{structure} (or
$\mathcal{C}$-structure) to indicate an element of a combinatorial
class $\mathcal{C}$. No particular assumption is ever made on the
nature of such structures, though classical examples tend to come from
discrete mathematics or theoretical computer science: words over some
finite alphabet, sequences, various flavors of trees, etc.

\subsection{Combinatorial Classes and Products}

Let $\mathcal{C}$ be some combinatorial class. The size of an object
$c\in\mathcal{C}$ will be noted $|c|$.  For any integer $n$, let
$c_n$ denote the number of objects in $\mathcal{C}$ with size $n$. The
ordinary (resp. exponential) \emph{generating function} for
$\mathcal{C}$ is
\begin{displaymath}
C(z) = \sum_{n} c_n z^n, \ \ \hbox{resp. } \tilde{C}(z) = \sum_{n}
c_n \frac{z^n}{n!};
\end{displaymath}
it is always assumed that the considered generating function has
positive radius of convergence $\rho$, \textit{i.e.} that
$\overline{\lim} c_n^{1/n}< \infty$ (resp., $\overline{\lim} (c_n/n!)^{1/n} < \infty$).

For $0< x<\rho$, the (normal, resp. exponential) \emph{Boltzmann
distribution over $\mathcal{C}$ for parameter $x$} is the probability
distribution defined, for any $c\in\mathcal{C}$, by
\begin{displaymath}
\pr_x (c) = \frac{x^{|c|}}{C(x)},\ \ \hbox{resp. } \pr_x(c) = \frac{x^{|c|}}{n! \tilde{C}(x)}.
\end{displaymath}

These distributions give positive probability to all objects in the
class, with the property that two objects with the same size have the
same probability. They are, of course, not the only probability
distributions with this property; their interest lies mostly in their
relationship with two common constructions in the combinatorial world:
the normal and labelled products.

Given two classes $\mathcal{A}$ and $\mathcal{B}$, their normal
product is just their Cartesian product $\mathcal{C}
= \mathcal{A}\times \mathcal{B}$, with size defined additively by
$|(a,b)| = |a|+|b|$ for $(a,b)\in\mathcal{A}\times\mathcal{B}$.

To define the labelled product, one has to assume that structures are
made up of both an unlabelled structure $c$ and a ``labelling'', a
permutation on a set whose is that of $c$. Think of a structure $c$ as
being formed of $|c|$ basic ``atoms'', each of which receives a
distinct label from $[[1,|c|]] = \{i\in \mathbb{Z}: 1\leq i\leq
|c|\}$. By a slight abuse of notation, we identify these atoms with
the integers $1$ to $|c|$, so that the labellings are just
permutations $\sigma\in\mathcal{S}_{|c|}$. For each unlabelled
structure $c$, the set of admissible permutations may be a strict
subset of $\mathcal{S}_{|c|}$.

Then, the labelled product of two labelled structures $(a,\sigma_a)$
and $(b,\sigma_b)$ is defined as $(a,b)$, with admissible labelings
obtained by taking all partitions of $[[1,|a|+|b|]]$ into two parts
$A$ and $B$ of respective sizes $|a|$ and $|b|$, and, for each
partition, taking the one permutation
$\sigma\in \mathcal{S}_{|a|+|b|}$ where all entries in $A$ are in the
same respective order as that of $\sigma_a$, and all entries in $B$
are in the same respective order as that of $\sigma_b$ (that is, if
$(x,y)\in A^2$, then $\sigma_a(x)< \sigma_b(y)$ iff
$\sigma(x)<\sigma(y)$, and similarly for $B$). As a result,
$(a,\sigma_a)$ and $(b,\sigma_b)$ have ${|a|+|b| \choose |a|}$
different structures in their labelled product. The labelled product
of two classes is defined as the set of all labelled products of
structures in the two original classes, with size again defined
additively.

The first important property is as follows: if two unlabelled
(resp., labelled) classes $\mathcal{A}$ and $\mathcal{B}$ have
generating functions $A(z)$ and $B(z)$ (resp., exponential generating
functions $\tilde{A}(z)$ and $\tilde{B}(z)$), then their normal
(resp. labelled) product has generating function $C(z)=A(z)B(z)$
(resp., $\tilde{C}(z)=\tilde{A}(z) \tilde{B}(z)$).

An immediate, and most useful, consequence, valid under both models,
is this: \emph{if $\mathcal{C}=\mathcal{A}\times \mathcal{B}$, then
taking the product of two independent $\mathcal{A}$-structure and
$\mathcal{B}$-structure, each following the Boltzmann distribution
with parameter $x$, results in a $\mathcal{C}$-structure under the
Boltzmann distribution with parameter $x$.} For labelled structures,
it is implied that one selects a uniform random set to define the
permutation in the product.

This ``independence under substructures'' property, in turn, has
interesting practical consequences, in that it makes it possible to
describe, for a number of classical combinatorial constructions,
systematic ways to produce efficient algorithms to sample from the
Boltzmann distribution for classes that are entirely described
(possibly in a recursive way) with them. This is the topic of
Section~\ref{sec:constructions}.

\subsection{The Boltzmann Method}

By a \emph{Boltzmann sampler} for a combinatorial class $\mathcal{C}$,
we mean a randomized algorithm $\Gamma\mathcal{C}$ that takes as input
a real parameter $x$ and outputs a random $\mathcal{C}$-structure
under the Boltzmann distribution with parameter $x$. Our overall goal
is to create efficient Boltzmann samplers for as many combinatorial
classes as possible.

Given a description of a combinatorial class $\mathcal{C}$ from which we
would like to obtain ``large random structures'' (of size $n$,
ideally), the Boltzmann method can be summarized as follows:
\begin{enumerate}
\item Find out if our class can be specified (up to a reasonably simple
size-preserving bijection) from the constructions in
Section~\ref{sec:constructions}. If not, try to extend the expressive
power of the method by adding new constructions. If this does not
work, maybe the Boltzmann method is not the best choice after all.
\item Use the techniques of the Purple Book~\cite{FlSe09} to locate the
``dominant'' singularities (those of smallest modulus, which govern
the asymptotics of the counting coefficients; since the generating
functions have nonnegative coefficients, at least one such singularity
lies on the positive real axis) of the generating functions in our
specification, and possibly an estimate of the value we should give
parameter $x$ to give expected size $n$ to $\mathcal{C}$-structures
under the Boltzmann distribution.
\item Compute approximations (to roughly $\Theta(\log n)$ digits) of the
values at $x$ of all involved generating functions, possibly using the
combinatorial oracle of Section~\ref{sec:gf_eval}.
\item Use the patterns in Section~\ref{sec:constructions}, together with our
specification, to write a Boltzmann sampling program.
\item Optionally, add a rejection scheme to obtain samples with size
in $[(1-\epsilon)n,(1+\epsilon)n]$, or even of exact size $n$ (more costly).
\end{enumerate}

Alternatively, step 2 can be replaced by experimentation using the other steps.

\subsection{Choosing the Parameter and Tuning for Size}

As said above, the Boltzmann model uses a real parameter that may be
chosen arbitrarily inside the radius of convergence of the generating
function for the class under consideration. This parameter $x$ governs
the distribution of sizes of the random structures. In typical
applications, one would like to obtain \emph{large} structures, that
is, there is an ideal value $n$ for sizes. We now briefly turn to the
question of picking an appropriate value of $x$ for a target $n$.

For a given $x$, the probability that a Boltzmann-distributed
$\mathcal{A}$-structure will have size $n$ is given by
\begin{displaymath}
p_n = p_n(x) = \frac{a_n x^{n}}{A(x)};
\end{displaymath}
multiplying by $n$, and summing over all values of $n$, we obtain
the \emph{expected} size
\begin{displaymath}
N = N_{\mathcal{A}}(x) = \frac{x A'(x)}{A(x)},
\end{displaymath}
where $A'$ is just the derivative of the generating function $A$
(which can be formally defined as $A'(x)=\sum_{n} n a_n x^{n-1}$; this
corresponds to the usual derivative inside the radius of convergence
of $A$).

Except in degenerate cases, $N_{\mathcal{A}}$ is a strictly increasing
and convex function on the interval $[0,\rho)$, and the equation
$N(x)=n$ has at most one solution; if, as is often the case, $A'(x)$
goes to infinity as $x$ goes to $\rho$, the equation has a unique
solution $x_n$ for each integer $n$. Setting $x$ to this value $x_n$
in the Boltzmann samplers from the previous section results in a
sampling algorithm that produces structures of expected size
$n$. Interestingly enough, the same equation also describes the value
of $x$ that maximizes the probability $p_n$ that the output structure
will have size $n$.

When the generating function is known exactly, one can solve for the
exact value of $x_n$. In many situations of interest, the generating
function is known only through an equation that it satisfies (this is
typically the case when the class is defined recursively, as described
in the next section). In this case, it is often possible to use the
techniques of analytic combinatorics~\cite{FlSe09} to derive precise
asymptotic information about the generating function and obtain a
precise estimate of $x_n$.

In some cases, an attractive alternative is to use the singularity
$x=\rho$; although it often implies an infinite expected size, a
simple adaptation of the sampling algorithms makes this a very viable
choice. This will be described in more detail in
Section~\ref{sec:size}.

\section{BASIC CONSTRUCTIONS AND BOLTZMANN SAMPLERS}
\label{sec:constructions}

In this section, we describe a number of constructions that can be
used to describe more complex classes from simpler ones, and, for each
construction, the corresponding combination algorithm that allows one
to build a Boltzmann sampler for the new class using
Boltzmann samplers for the classes involved in the description. The
constructions described here allow one to describe combinatorial
classes that are close to those of the theory of combinatorial
species~\cite{BeLaLe98}.

When nothing is specified, these constructions apply to both labelled
and unlabelled structures; in the labelled case, it is silently
assumed that one performs a label redistribution as in the case of the
labelled product.

The initial constructions were described
in~\cite{DuFlLoSc04}; later additions are credited
individually.

In all cases, the construction is translated into an expression for
the generating function of the new class in terms of the previous one;
this in turns gives a simple construction for the sampling algorithm,
where ``substructures'' are independent. Many algorithms can be
expressed in the form ``Draw integer $k$ from discrete distribution
$\mu(x)$, then let $\gamma$ receive the concatenation of $k$
independent calls to generator $\Gamma\mathcal{A}(x)$''; we abbreviate
this as
\begin{displaymath}
\gamma \leftarrow \left[ \Gamma\mu(x) \implies \Gamma\mathcal{A}(x)\right].
\end{displaymath}

We also use samplers for a few standard distributions: Bernoulli with
success probability $x$ ($\hbox{Bern}(x)$), geometric (with support
$\mathbb{N}$) with parameter $x$ ($\hbox{Geom}(x)$), and Poisson with
rate $x$ ($\hbox{Poiss}(x)$). A subscript condition on these samplers
means a conditioning on the output, which can be achieved by rejecting
outputs until the condition is met.

\subsection{Finite Sets}

Finite (typically small) sets do not require an elaborate theory to
produce sampling algorithms, and are included to serve as elementary
bricks for more complex constructions.

One typically defines an ``empty structure'' class $\mathcal{E}$,
containing a single structure of size zero that we denote as $\mathbf{1}$,
and an ``atom'' class $\mathcal{Z}$, containing a single structure $Z$
of size $1$. Occasionally, one may use a number of different atom ``types'',
which will then be written $Z_a,Z_b$, and so on.

\subsection{Disjoint Union}

The most basic construction is that of \emph{disjoint union}: if
$\mathcal{A}$ and $\mathcal{B}$ are disjoint classes, then their union
$\mathcal{C}=\mathcal{A}\cup \mathcal{B}$ (with size inherited from
the original class) is a new class, whose generating function is simply
$C(z)=A(z)+B(z)$. 

\begin{algorithm}
\begin{algorithmic}
\STATE \textbf{Algorithm} $\Gamma[\mathcal{A}\cup \mathcal{B}]$

\IF{$\hbox{Bern}(A(x)/(A(x)+B(x)))$}
  \STATE Return $\Gamma\mathcal{A}(x)$
\ELSE
  \STATE Return $\Gamma\mathcal{B}(x)$
\ENDIF
\end{algorithmic}
\end{algorithm}

\subsection{Product}

The product construction, being of fundamental importance, has been
described in the previous section.

\begin{algorithm}
\begin{algorithmic}
\STATE \textbf{Algorithm} $\Gamma[\mathcal{A}\times \mathcal{B}]$

\STATE Return $(\Gamma\mathcal{A}(x),\Gamma\mathcal{B}(x))$
\end{algorithmic}
\end{algorithm}

\subsection{Sequence}

If $\mathcal{A}$ is a class with no structures of size $0$,
$\mathcal{C}=\hbox{Seq}(\mathcal{A})$ is the set of sequences
$(A_1,\dots,A_k)$, for arbitrary $k\geq 0$, with $A_i\in\mathcal{A}$, and
size defined additively by
\begin{displaymath}
        |(A_1,\dots,A_k)| = |A_1| + \dots + |A_k|.
\end{displaymath}

The generating function for $\mathcal{C}$ is the \emph{pseudo-inverse} of that of $\mathcal{A}$,
\begin{displaymath}
C(z) = \frac{1}{1-A(z)},
\end{displaymath}
and the corresponding Boltzmann sampler for $\mathcal{C}$ is as
follows:

\begin{algorithm}
\begin{algorithmic}
\STATE \textbf{Algorithm} $\Gamma\hbox{Seq}(\mathcal{A})$

\STATE $\gamma \leftarrow \left[ \hbox{Geom}(A(x)) \implies \Gamma\mathcal{A}(x)\right]$
\STATE Return $\gamma$
\end{algorithmic}
\end{algorithm}

\subsection{Cycle (labelled)}

If $\mathcal{A}$ is a class with no structures of size $0$,
$\mathcal{C}=\hbox{Cycle}(\mathcal{A})$ is the set of \emph{cycles} of
$\mathcal{A}$-structures, that is, sequences defined up to a circular
permutation of the component $\mathcal{A}$-structures.

Working with labelled structures means that each sequence of $k$
structures has exactly $k-1$ other structures that correspond to the
same cycle. Consequently, the generating function is
\begin{displaymath}
\tilde{C}(z) = \sum_{k\geq 1} \frac{\tilde{A}^k(z)}{k} = - \log(1-\tilde{A}(z)),
\end{displaymath}
and the corresponding Boltzmann sampler uses the ``logarithmic''
distribution $\mu_{x}(k) = \frac{x^k}{k |\log(1-x)|}$ ($k\geq 1$),
provided by sampler $\hbox{Loga}()$:

\begin{algorithm}
\begin{algorithmic}
\STATE \textbf{Algorithm} $\Gamma\hbox{Cycle}(\mathcal{A})$

\STATE $\gamma \leftarrow \left[ \hbox{Loga}(\tilde{A}(x)) \implies \Gamma\mathcal{A}(x) \right]$
\STATE Return $\gamma$
\end{algorithmic}
\end{algorithm}

\subsection{Set}

If $\mathcal{A}$ is a labelled class with no structures of size $0$,
$\mathcal{C}=\hbox{Set}(\mathcal{A})$ is the class of \emph{sets} of
$\mathcal{A}$-structures, that is, (possibly empty) sequences up to an
arbitrary permutation of component structures.

The corresponding generating function is
\begin{displaymath}
\tilde{C}(z) = \sum_{k\geq 0} \frac{\tilde{A}^k(z)}{k!} = \exp(\tilde{A}(z)),
\end{displaymath}
and the distribution for the number of components is
the \emph{Poisson} distribution:

\begin{algorithm}
\begin{algorithmic}
\STATE \textbf{Algorithm} $\Gamma[\hbox{Set}(\mathcal{A})]$
\STATE $\gamma \leftarrow \left[ \hbox{Poiss}(\tilde{A}(x)) \implies \Gamma\mathcal{A}(x) \right]$
\STATE Return $\gamma$
\end{algorithmic}
\end{algorithm}

A set construction for unlabelled structures was introduced
in~\cite{FlFuPi07}. It is based on the multiset construction described
next, and requires more elaborate manipulations.

\subsection{Multiset (unlabelled)}

Boltzmann samplers for the Multiset construction were introduced
in~\cite{FlFuPi07}.

If $\mathcal{A}$ is an unlabelled class with no structures of size
$0$, $\mathcal{C} = \hbox{MSet}(\mathcal{A})$ is the class of
all \emph{multisets} of $\mathcal{A}$-structures - sets with possible
repetitions.

The corresponding generating function is
\begin{displaymath}
C(z) = \exp\left( \sum_{k\geq 1} \frac{1}{k}A(z^k)\right).
\end{displaymath}

The Boltzmann sampler uses a $\hbox{MaxIndex}(A,x)$ subroutine, which
samples from the discrete distribution defined by
\begin{displaymath}
\mathbb{P}_{A,x}(K\leq k) = \frac{1}{C(x)}\prod_{j\leq k} \exp\left(\frac{1}{j} A(x^j)\right).
\end{displaymath}

The Boltzmann sampler itself is then as follows:

\begin{algorithm}
\begin{algorithmic}
\STATE \textbf{Algorithm} $\Gamma\hbox{MSet}(\mathcal{A})$
\STATE $\gamma\leftarrow \emptyset$
\STATE $k\leftarrow \hbox{MaxIndex}(A,x)$
\FOR{$j$ from 1 to $k-1$}
  \STATE $\gamma \leftarrow \gamma \cup \left[ \hbox{Poiss}(A(x^j)/j) \implies j\hbox{\ copies of }\Gamma\mathcal{A}(x^j)\right]$
\ENDFOR
\STATE $\gamma \leftarrow \gamma\cup \left[ \hbox{Poiss}_{\geq 1}(A(x^k)/k) \implies k\hbox{\ copies of }\Gamma\mathcal{A}(x^k)\right]$.
\STATE Return $\gamma$
\end{algorithmic}
\end{algorithm}

(It should be noted that $k$ in the above algorithm is \emph{not} the
maximum number of repetitions of a structure in the output, but only a
lower bound: all calls to the sampler $\Gamma\mathcal{A}$ are
independent, so that a structure may be output more than once and
obtain larger multiplicity.)

Note that, in contrast to the previous constructions, the value $C(z)$
is expressed not in terms of the value $A(z)$, but of the values of
$A$ for a whole geometric sequence of values.

\subsection{Recursive Constructions}

All of the above constructions can be
used \emph{recursively}, \textit{i.e.} a class $\mathcal{C}$ can be
defined using one of these constructions on a class that is itself
(ultimately) defined in terms of $\mathcal{C}$ itself. Some care must
be taken to avoid circular definitions: recursive specifications
define structures from smaller structures, possibly of the same type,
but not from themselves. Thus, one can define a class $\mathcal{P}$ by
$\mathcal{P} = \mathcal{Z}\times \hbox{Seq}(\mathcal{P})$ (this
defines plane trees: a plane tree is composed of a root having an
ordered sequence (possibly empty) of children, each the root of a
plane tree), but a specification such as $\mathcal{P}
= \mathcal{A}\times \mathcal{P}$ would be invalid if class
$\mathcal{A}$ contains structures of size $0$ (it would then attempt
to create an infinite number of $\mathcal{P}$-structures of size $0$).

Subject to this ``well-foundedness'' condition, all the previous
constructions can be used recursively - and indeed, in most
applications of interest recursivity is used. Whenever it is the case,
the Boltzmann samplers derived from the previous subsections become
recursive algorithms, for which termination can only be guaranteed
with probability $1$ (and in finite expected time; see
Section~\ref{sec:complexities}).

An effective characterization of this ``well-foundedness'' condition
is given in~\cite{PiSaSo08}, for specifications that outright forbid
structures of size $0$.

\subsection{Ordered Structures and Differential Operators (labelled)}

The constructions described in this subsection appear
in~\cite{RoSo09,BoRoSo11}.

The derivative $\alpha'$ of a labelled combinatorial structure
$\alpha$ is obtained by replacing the atom in $\alpha$ having the
largest label with a ``hole'' - this hole holds the place of an atom, but
does not contribute to size and does not get a label. Thus, the
derivative of a structure of size $n$ has size $n-1$. The derivative
of a combinatorial class $\mathcal{A}$ is, of course, the class
$\mathcal{C} = \mathcal{A}' = \{\alpha': \alpha\in\mathcal{A}\}$ of
all derivatives of $\mathcal{A}$-structures. The corresponding
(exponential) generating functions are related by
\begin{eqnarray*}
\tilde{C}(z) & = & \tilde{A}'(z),\\
\tilde{A}(z) & = & a_0 + \int_{0}^{z} \tilde{C}(z) dz,
\end{eqnarray*}
where $a_0$ is the number of $\mathcal{A}$-structures of size $0$.

Derivative classes can be used in recursive constructions, under
suitable ``well-foundedness'' conditions (see~\cite{Bo10,BoRoSo11} for
details), to define a class as the solution to a symbolic differential
equation. In very rough terms, this corresponds to imposing order
conditions on labels. A simple example is provided by the class $\mathcal{T}$
of \emph{decreasing (labelled) binary trees}, that is, labelled binary
plane trees where each node (atom) is required to have a larger label
than each of its children: the equation reads
\begin{eqnarray*}
\mathcal{T}' & = & \epsilon \cup \mathcal{T}\times \mathcal{T}\\
\mathcal{T}_0 & = & \emptyset.
\end{eqnarray*}
and should be understood as this: the largest label in a decreasing
binary tree has to be at the root, so its derivative will either be a
unique object of size zero or equivalent (after relabelling) to a pair
of binary trees, each of which has to be decreasing.

Bodini \textit{et al.}~\cite{BoRoSo11} describe a generic Boltzmann sampler for a class defined
by a first-order differential operator $\mathcal{A}'
= \mathcal{F}(\mathcal{Z},\mathcal{A})$, provided one has a Boltzmann
sampler for the class $\mathcal{F}(\mathcal{Z},\mathcal{A})$ (that is,
$\mathcal{F}$ is defined in terms of other classical constructions,
and the whole sampler will necessarily be recursive). Like the sampler
for multisets, it requires a change of the parameter - here, a random
change - for each recursive call.

Given the generating function $A$ and a parameter $0<x_0<\rho_A$, one
defines a probability density on the interval $[0,1]$ by
\begin{displaymath}
h_{x_0,A}(u) = \frac{x_0 A'(ux_0)}{A(x_0)-A(0)};
\end{displaymath}
if $U$ is a random variable following this distribution, $Ux_0$ can be
interpreted as the result of picking a random point (according to
Lebesgue measure) in the domain $0<y<A(x), 0<x<x_0$, and keeping the
abscissa $x$.

With this definition, the Boltzmann sampler is as follows:
\begin{algorithm}
\begin{algorithmic}
\STATE \textbf{Algorithm} $\Gamma\mathcal{A}$ from $\Gamma\mathcal{A}'$, $\mathcal{A}'= \mathcal{F}(\mathcal{Z},\mathcal{A})$

\IF{$\hbox{Bern}(A(0)/A(x))$}
  \STATE Return a uniform object from $\mathcal{A}_0$
\ELSE
  \STATE Draw $U\in[0,1]$ following density $h_{x,A}$
  \STATE $f\leftarrow \Gamma\mathcal{F}[\mathcal{Z},\mathcal{A}](Ux)$
  \STATE Return object $(Z,f)$ with atom $Z$ having the largest label
\ENDIF
\end{algorithmic}
\end{algorithm}

\subsection{\hspace{.7em}Multivariate Models}

So far, we have only considered generating functions with a single
variable, which ``counts'' for the size of the structures; that is,
each structure $\gamma$ in the class contributes a single term
$z^{|\gamma|}$ or $z^{|\gamma|}/\gamma!$ to the generating function.

Given a combinatorial class $\mathcal{C}$, one can define an arbitrary
number of statistics $s_i: \mathcal{C}\rightarrow \mathbb{N}$ ($1\leq
i\leq k$), and the corresponding multivariate generating function
$C(z,u_1,\dots,u_k)$ (first as a multivariate formal power series,
then as an analytic function) by changing the contribution of each
structure $c\in\mathcal{C}$ to $z^{|c|} \prod_{1\leq i\leq k}
u_i^{s_i(c)}$ (for the rest of this subsection, we assume unlabelled
structures), and consequently, a Boltzmann distribution for any tuple
$(x,u_1,\dots,u_k)$ of positive real variables lying inside the
convergence domain of the generating function. 

When the considered statistics are transmitted additively under the
constructions described in this section (say, if there are several
types of ``atoms'', and statistic $s_i$ counts the number of atoms of
type $i$), the Boltzmann samplers can be adapted to this generalized
model.

This area has not been explored as extensively as others, probably
because a general theory would involve additional technical
details. Bodini and Ponty~\cite{BoPo10} use it to sample from context-free languages
with a nonuniform distribution where the frequency of letters is
artificially skewed, with an application to ``Tetris tesselations''
(perfect tilings of a rectangular region with pentominoes) where each
piece has the same frequency.

\section{ALGORITHM COMPLEXITIES}
\label{sec:complexities}

Each of the individual algorithms in Section~\ref{sec:constructions}
has low overhead complexity, but they tend to make possibly unbounded
numbers of calls to other algorithms. The general theorem below is a
compilation of results from \cite{DuFlLoSc04} and other papers that
extend the expressive power of ``specifiable'' classes.

\begin{theorem}
Let $\mathcal{C}$ denote a (labelled or unlabelled) class that can be
entirely specified, in a possibly recursive way, with the
constructions of Section~\ref{sec:constructions}, and let $0<x<\rho$
be any positive real lying inside the convergence domain for the
generating function of $\mathcal{C}$. Then, assuming an oracle that
provides values of the relevant generating functions at real values,
the algorithm $\Gamma\mathcal{C}$ compiled from the specification by
the patterns of Section~\ref{sec:constructions} terminates with
probability $1$ and in finite expected time, outputs a random
$\mathcal{C}$-structure under the Boltzmann distribution with
parameter $x$, and uses a number of real arithmetic operations that
is \emph{linear} in the size of the output.
\end{theorem}

\section{APPROXIMATE AND EXACT SIZE SAMPLERS}
\label{sec:size}

The user of random generation algorithm is often used to \emph{exact}
random samplers (algorithms that take $n$ as input, and output a
uniform random element of the subclass $\mathcal{C}_n$), or
perhaps \emph{approximate-size} samplers (algorithms that take two
integers $n<N$ as input, and output a random element with size in
$[n,N]$ with equal probability for any two elements with the same
size). Both types can be obtained by adding a rejection mechanism to
Boltzmann samplers, at overall costs that depend on the distribution
of sizes under the Boltzmann model.

We denote $\mu_1(x)$ for the expected
size, $\mu_2(x)$ for the expected squared size, and $\sigma(x)$ for
the standard deviation on size, all as functions of parameter $x$;
analytically,
\begin{eqnarray*}
\mu_1(x) & = & \frac{x C'(x)}{C(x)}\\
\mu_2(x) & = & \frac{x C'(x) + x^2 C''(x)}{C(x)}\\
\sigma(x) & = & \sqrt{\mu_2(x) - \mu_1^2(x)}
\end{eqnarray*}

\subsection{Approximate Size Samplers}

Assume we are given a target size $n$, some tolerance $\epsilon>0$,
and the value $x_n$ of the parameter that ensures that the expected
size of structures is $n$. We obtain an approximate size sampler with
acceptable sizes in $I = I(n,\epsilon) =
((1-\epsilon)n,(1+\epsilon)n)$ by repeatedly using the Boltzmann
sampler $\Gamma\mathcal{C}(x_n)$, until a structure whose size lies in
$I$ appears. 

Since the cost of each call to the Boltzmann sampler is linear in the
size of the output by Theorem 1, the expected cost of this approximate
size sampler is asymptotic to $n$ times the expected number of calls
to the Boltzmann sampler. 

In favorable situations, described as ``lumpy'' distributions
in~\cite{DuFlLoSc04} and characterized by
$\sigma(x)/\mu_1(x) \rightarrow 0$ as $x\rightarrow \rho^{-}$, this
expected number of calls goes to $1$ as $n$ (the target size) goes to
infinity -- the probability that the first call will yield a structure
whose size is in $I$ is asymptotically $1$. More precise information
on the distribution of sizes produced by such an approximate-size
sampler can be obtained through the asymptotics of the generating
function. Typically, for ``lumpy'' distributions, this size is
concentrated around the expected size, and shorter intervals of length
$o(n)$ could be used without altering the theoretical results.

In less favorable situations, this success probability goes to a
finite positive constant as $n$ goes to infinity, so that the expected
cost of the approximate size sampler is still asymptotically
linear. In many cases, a size tolerance of, say, $5\%$ around the
target size, at the cost of a constant number of rejections, is quite
acceptable.

In some even less favorable situations, it may be necessary to change
the class by using ``pointing'' (a combinatorial operation close to
derivation, corresponding to distinguishing a single atom in the
structure) a finite number of times before one gets to the situation
described above.

\subsection{Exact Size Samplers}

Setting $\epsilon=1/n$ in the approximate size samplers of the
previous subsection, results in exact size samplers. It should be
noted, however, that in most cases, the success probability (the
probability, with parameter $x_n$, of obtaining a structure of size
exactly $n$) is only of order $\Theta(1/n)$, which results in an
expected complexity $\Theta(n^2)$ for the exact size sampler. 

\subsection{Singular Samplers}

Whenever the generating function is finite at its dominant singularity
$\rho$, one can define a Boltzmann distribution for $x=\rho$, and the
Boltzmann samplers can be used with parameter $\rho$. This is
typically (though not universally) true with recursive specifications,
and the most frequent case is for the generating function to have a
``square root-type'' singularity, \textit{i.e.} $C(z)$ has a singular
expansion of the form $C(z) = C(\rho) + a (1-z/\rho)^{1/2} +
o((1-z/\rho)^{1/2})$ as $z$ approaches $\rho$.

In such cases, the expected size for the singular Boltzmann model is
infinite. While this offers the best chances of success for the
approximate size samplers, it also implies that the expected cost of
the same approximate size sampler is infinite, which is unacceptable.

However, on closer examination, this infinite expectation only comes
from those (rare: the probability is $\Theta(1/\sqrt{n})$) runs of the
sampler where the output size is much larger than the target $n$. A
simple modification of the Boltzmann samplers can thus avoid this
higher cost, by keeping track of the number of atoms generated so far
and aborting the Boltzmann sampler as soon as the total becomes larger
than $n$; with this modification, expected costs for a square root
singularity become $\Theta(n)$ for approximate size with finite
$\epsilon$, and $\Theta(n^{3/2})$ for exact size.

\section{GENERATING FUNCTION EVALUATION AND PRECISION}
\label{sec:gf_eval}

One of the key points of the Boltzmann method, when compared to the
recursive method, is that enumeration sequences are replaced
by \emph{generating function evaluations}. In many of the
constructions of Section~\ref{sec:constructions}, each involved generating
function needs to be evaluated for the same value of its variable; in
a few of them, some have to be evaluated for a deterministic or random
sequence of values.

In some cases, the generating functions have closed forms and this
evaluation does not lead to special complications, but recursive
specifications lead to generating functions that are determined by
equations, and the question of determining a good approximation of the
required values becomes more troublesome.

Pivoteau, Salvy and Soria~\cite{PiSaSo08} provide an efficient solution to this problem, at least 
for the basic constructions of sums, products, sequences, cycles and
sets. The natural idea of iterating the equations provides only slow
convergence; the preferred method is based on Newton iteration, which
ensures quadratic convergence (asymptotically, distance to the exact
solution is squared by each iteration). For a given value of the
variable, the generating function equations typically have several
real solutions, only one of which corresponds to the generating
function that is the ``true'' solution; an important result in the
above-mentioned paper is that, thanks to the existence of a
combinatorial equivalent to the Newton iteration, convergence to the
``true'' value is ensured.

Also, note that, when applying Newton iteration for generating
function evaluation, convergence is significantly faster the further
$x$ is from the dominant singularity $\rho$. While this is good for
constructions such as multisets or ordered structures that require the
use of sequences of values (for sequences of values of the variable
that quickly decrease to zero), it is conversely bad news for
classical applications that require very large structures, since this
means taking values of $x$ that are very close to
$\rho$. Nevertheless, the experimental results reported
in~\cite{PiSaSo08}, even for very complex specifications, remain
within reasonable bounds (the computation time for the oracles of
specifications implying 500 equations, for values of the parameter
leading to expected structure sizes in the tens of thousands, are of
the order of a minute).

Another question that arises naturally is that of the influence of
approximations on the final distribution of random samples. Even
assuming a ``perfect'' source of randomness for the simulations, a
small error in a generating function value that is used repeatedly by
a sampling algorithm might result in a significantly biased
distribution of the final samples. 

To give an example, suppose a specification involves a disjoint union
$\mathcal{C}=\mathcal{A}\cup \mathcal{B}$, and each of $\mathcal{A}$
and $\mathcal{B}$ is defined recursively using $\mathcal{C}$ - this is
not an artificial assumption. Assume that, for the value $x$ of the
parameter, $A(x)$ is slightly overestimated, and $B(x)$ is slightly
underestimated. In this case, each time the sampler for $\mathcal{C}$
is used, it will have a tendency to switch to $\mathcal{A}$ more often
than it ideally should; this will result in a distribution that is
biased in favor of $\mathcal{C}$-structures that often use
$\mathcal{A}$-structures as components.

One possible solution~\cite{Du11} to estimate, and possibly correct,
this bias, is to design Boltzmann samplers that not only output a
random structure, but also, for each real-valued constant $A\simeq
A(x)$ used in the sampling, a ``safety interval'' $[A^{-},A^{+}]$,
with $A^{-}<A<A^{+}$, with a precise meaning of ``if the sampler had
been run with any value in $[A^{-},A^{+}]$ instead of $A$, the result
of the whole computation would have been the same''. This is done
by studying the small number of discrete distributions one really
needs to sample from.

From such ``safety interval'' samplers, one can derive both a
practical and a theoretical result:
\begin{itemize}
\item An estimate of the quality of approximation one should have on each
involved constant, such that the whole Boltzmann sampler is very
unlikely to output any safety interval that does \emph{not} contain the
true value of the corresponding generating function. Not surprisingly,
$\Theta(\log n)$ bits are enough when the expected size is $n$.
\item A (still hypothetical) construction for a \emph{truly exact} Boltzmann
sampler, if one assumes stronger oracles than those provided
by \cite{PiSaSo08}. If one assumes oracles that give both an upper an
a lower bound for each involved constant, and that the oracles can be
called repetitively to decrease the difference between these bounds
(say, by a factor of 2 with every iteration), then the ``safety
interval'' approximate Boltzmann samplers can be modified into exact
Boltzmann samplers which will call the oracles a small (logarithmic)
number of times on average.
\end{itemize}

\section{NOT-QUITE BOLTZMANN SAMPLERS}
\label{sec:not-quite}

Boltzmann samplers share two important and useful properties for the
user interested in practical random sampling of large structures: (1)
they output structures of random size, with the guarantee
that \emph{any two structures of the same size have the same
probability of being obtained}, and (2) they are \emph{``stable''
under the constructions} of Section~\ref{sec:constructions}.

While property (1) is the essential one for direct practical
applications, making it possible to use, say, rejection to transform
Boltzmann samplers into exact- or approximate-size samplers, property
(2) is the one responsible for the wide applicability of the
method. Nevertheless, sometimes property (1) alone can be obtained
while keeping algorithms of low complexity, typically by using a
``nonstandard'' final construction using Boltzmann samplers as
subroutines.

A good example of this is the \emph{Hadamard product} of two
combinatorial class. If $\mathcal{A}$ and $\mathcal{B}$ are two
arbitrary (unlabelled) combinatorial classes, their Hadamard product
$\mathcal{C}= \mathcal{A}\odot\mathcal{B}$ is the subset of
$\mathcal{A}\times\mathcal{B}$ that only contains pairs $c=(a,b)$ with
the same size (with $|c|$ defined to be $|a|=|b|$ instead of the sum
for the classical product). The corresponding generating function is
none other than the Hadamard product of generating functions,
\begin{displaymath}
C(z) = A\odot B(z) = \sum_{n} a_n b_n z^n,
\end{displaymath}
with radius of convergence at least the product of radii of $A(z)$ and
$B(z)$.

A real Boltzmann sampler for $\mathcal{C}$ can be written easily: one
simply checks that, if $x=x_A x_B$ with $x_A<\rho_A$ and $x_B<\rho_B$,
the algorithm
\begin{algorithmic}
\REPEAT
  \STATE $\alpha \leftarrow \Gamma\mathcal{A}(x_A)$
  \STATE $\beta \leftarrow \Gamma\mathcal{B}(x_B)$
\UNTIL{$|\alpha| = |\beta|$}
\STATE Return $(\alpha,\beta)$
\end{algorithmic}
terminates with probability 1 and in finite expected time, and outputs
a $\mathcal{C}$ structure under the Boltzmann distribution with
parameter $x$. It is, however, inefficient: each iteration has
success probability $C(x)/(A(x_A) B(x_B))$, which can be very small
when $x$ is close to the dominant singularity.

Bodini \textit{et al.}~\cite{BoGaRo10} use the classical Birthday paradox to devise a more
efficient algorithm that preserves property (1), though not the whole
Boltzmann distribution: simply alternate (either deterministically or
randomly) between $\Gamma\mathcal{A}(x_A)$ and
$\Gamma\mathcal{B}(x_B)$, retaining only the first obtained structure
of each class and size, until a pair with the same size can be formed.

\section{CONCLUSION}

Boltzmann samplers are an attractive class of random generation
algorithms for classes of combinatorial structures that lend
themselves to combinatorial decompositions, allowing for fast
(linear-time, or quasi-linear-time) generation of structures with size
in the millions for simple classes, and well into the tens of
thousands for complex classes. 


\bibliographystyle{plain}
\bibliography{minimalbib}

\begin{thebibliography}{10}

\bibitem{BeLaLe98}
Fran{\c c}ois Bergeron, Gilbert Labelle, and Pierre Leroux.
\newblock {\em Combinatorial species and tree-like structures}.
\newblock Number~67 in Encyclopedia of Mathematics and its Applications.
  Cambridge University Press, Cambridge, 1998.

\bibitem{Bo10}
Olivier Bodini.
\newblock {\em Autour de la g\'en\'eration al\'eatoire sous mod\`ele de
  {B}oltzmann}.
\newblock Habilitation memoir, Universit\'e Pierre et Marie Curie, 2010.
\newblock in French.

\bibitem{BoGaRo10}
Olivier Bodini, Dani\`ele Gardy, and Olivier Roussel.
\newblock Boys-and-girls birthdays and {H}adamard products.
\newblock In {\em Proceedings of Lattice Paths Combinatorics and Applications},
  2010.

\bibitem{BoPo10}
Olivier Bodini and Yann Ponty.
\newblock Multi-dimensional {B}oltzmann sampling of languages.
\newblock In {\em Proceedings of the 21st International Meeting on
  Probabilistic, Combinatorial, and Asymptotic Methods in the Analysis of
  Algorithms (AofA'10)}, pages 49--64. DMTCS Proceedings, 2010.

\bibitem{BoRoSo11}
Olivier Bodini, Olivier Roussel, and Mich\`ele Soria.
\newblock {B}oltzmann samplers for first order differential specifications.
\newblock {\em To appear in \emph{Discrete Applied Mathematics}}, 2011.

\bibitem{Du11}
Philippe Duchon.
\newblock Exact {B}oltzmann samplers from adaptive approximate oracles.
\newblock In preparation, 2011.

\bibitem{DuFlLoSc02}
Philippe Duchon, Philippe Flajolet, Guy Louchard, and Gilles Schaeffer.
\newblock Random sampling from {B}oltzmann principles.
\newblock In P.~Widmayer, F.~Triguero, R.~Morales, M.~Hennessy, S.~Eidenbenz,
  and R.~Conejo, editors, {\em Proceedings of the 29th ICALP}, volume 2380 of
  {\em Lecture Notes in Computer Science}, pages 501--513. Springer, 2002.

\bibitem{DuFlLoSc04}
Philippe Duchon, Philippe Flajolet, Guy Louchard, and Gilles Schaeffer.
\newblock Boltzmann samplers for the random generation of combinatorial
  structures.
\newblock {\em Combinatorics, Probability and Computing}, 13(4-5):577--625,
  2004.

\bibitem{FlFuPi07}
Philippe Flajolet, \'Eric Fusy, and Carine Pivoteau.
\newblock Boltzmann sampling of unlabeled structures.
\newblock In {\em Proceedings of ANALCO 07}, pages 201--211, New Orleans, 2007.

\bibitem{FlSe09}
Philippe Flajolet and Robert Sedgewick.
\newblock {\em Analytic combinatorics}.
\newblock Cambridge University Press, Cambridge, 2009.

\bibitem{FlZiVC94}
Philippe Flajolet, Paul Zimmermann, and Bernard Van~Cutsem.
\newblock A calculus for the random generation of labelled combinatorial
  structures.
\newblock {\em Theor. Comput. Sci.}, 132(2):1--35, 1994.

\bibitem{PiSaSo08}
Carine Pivoteau, Bruno Salvy, and Mich\`Ele Soria.
\newblock Boltzmann oracle for combinatorial systems.
\newblock In {\em Algorithms, Trees, Combinatorics and Probabilities}, pages
  475 -- 488. Discrete Mathematics and Theoretical Computer Science, 2008.
\newblock Proceedings of the Fifth Colloquium on Mathematics and Computer
  Science. Blaubeuren, Germany. September 22-26, 2008.

\bibitem{ProppWilsonCFTP}
James~Gary Propp and David~Bruce Wilson.
\newblock How to get a perfectly random sample from a generic markov chain and
  generate a random spanning tree of a directed graph.
\newblock {\em J. Algorithms}, 27(2):170--217, 1998.

\bibitem{Re85}
J.-L. R\'emy.
\newblock Un proc\'ed\'e it\'eratif de d\'enombrement d'arbres binaires et son
  application \`a leur g\'en\'eration al\'eatoire.
\newblock {\em RAIRO Theoretical Informatics and Applications}, 27(2):179--195,
  1985.

\bibitem{RoSo09}
Olivier Roussel and Mich\`ele Soria.
\newblock {B}oltzmann sampling of ordered structures.
\newblock In {\em Proceedings of LAGOS 09}, 2009.
\newblock 9 pages.

\end{thebibliography}

\section*{AUTHOR BIOGRAPHY}

\textbf{PHILIPPE DUCHON} is a Professor in the Computer Science department
of Universit\'e Bordeaux 1. Before joining the university, he was an
Associate Professor in the ENSEIRB engineering school. He is a former
student of \'Ecole Normale Sup\'erieure and received his Ph.D. in
Computer Science from Universit\'e Bordeaux 1 in 1998. His research
interests are in all areas of enumerative combinatorics, random
generation algorithms, and the analysis of algorithms for distributed
systems. His email address
is \href{mailto://duchon@labri.fr}{duchon@labri.fr}.

\end{document}